\documentclass[prb,preprint]{revtex4-1} 
\usepackage{graphicx}
\usepackage{amsmath}
\usepackage{amssymb}
\usepackage{hyperref}
\usepackage{caption}
\usepackage{subcaption}

\usepackage{tikz}
\usepackage{epstopdf}

\DeclareMathOperator\arccosh{arccosh}
\begin{document}

\title{Delayed Rebounds in the Two-Ball Bounce Problem}

\author{Sean P.~Bartz}
\email{sean.bartz@indstate.edu}
\affiliation{Dept.~of Chemistry and Physics, Indiana State University, Terre Haute, IN 47809}

\date{November 5, 2020}

\begin{abstract}
In the classroom demonstration of a tennis ball dropped on top of a basketball, the surprisingly high bounce of the tennis ball is typically explained using momentum conservation for elastic collisions, with the basketball-floor collision treated as independent from the collision between the two balls. This textbook explanation is extended to inelastic collisions by including a coefficient of restitution. This independent contact model (ICM), as reviewed in this paper, is accurate for a wide variety of cases, even when the collisions are not truly independent. However, it is easy to explore situations that are not explained by the ICM, such as swapping the tennis ball for a ping-pong ball.
In this paper, we study the conditions that lead to a ``delayed rebound effect," a small first bounce followed by a higher second bounce, using techniques accessible to an undergraduate student. The dynamical model is based on the familiar solution of the damped harmonic oscillator. We focus on making the equations of motion dimensionless for numerical simulation, and reducing the number of parameters and initial conditions to emphasize universal behavior. The delayed rebound effect is found for a range of parameters, most commonly in cases where the first bounce is lower than the ICM prediction.

\end{abstract}
\maketitle

\section{Introduction}
In a common classroom demonstration of linear momentum conservation, a tennis ball is held above a basketball, and the two are simultaneously dropped to the floor. 
The tennis ball rebounds much higher than the drop height, surprising and exciting students. 
Textbook explanations suggest that when the upper ball is much less massive than the lower ball, it rebounds at  three times the impact speed,  bouncing to nine times the initial drop height. \cite{mellen_superball_1968, harter_velocity_1971, herrmann_simple_1981}

An interested student may wonder about the accuracy of this prediction, as well as its applicability to other combinations of sports balls. 
For instance, replacing the tennis ball with a ping-pong ball more closely matches the textbook assumptions, suggesting it should bounce higher than the tennis ball. 
Instead, the ping-pong ball often stays close to the basketball on the first bounce. 
However, with the balls carefully aligned, the small first bounce is followed by a noticeably higher second bounce. 
This work shows that this ``delayed rebound effect" is robust, and is explained with a simple force model.

The classic justification for the high bounce of the tennis ball assumes that the basketball-floor collision is independent of the basketball-tennis ball collision. 
The simplifying assumption of this independent contact model (ICM) does not withstand  close inspection; the lower ball often remains in contact with the floor when the two balls first make contact.\cite{cross_vertical_2007} 
Interestingly, ICM predictions are fairly accurate in many cases when the collisions are not truly independent. \cite{berdeni_two-ball_2015}
However, when the initial separation between the two balls is small, the final velocities differ greatly from the ICM prediction, particularly when the balls are allowed to bounce more than once. 

General study of this problem seems to require consideration of an unwieldy number of parameters: the radii and masses of both balls, their elastic and dissipative constants, and the initial drop height and separation between the two balls. In this paper, we show how to reduce this set of eight parameters and two initial conditions to two parameters and a single initial condition. This type of generalization is  important  for applying computational techniques to understand universal behavior.



\section{Independent Contact Model\label{ICM}}

The textbook solution to the two-ball drop problem assumes independent, instantaneous collisions between the balls and the floor. 
We review this independent contact model here as a basis of comparison for the more-realistic dynamic model, and to introduce notation.
We focus on the maximum bounce height of the upper ball, as this is the effect most easily seen in the classroom demonstration. 

\begin{figure}
\begin{center}

\begin{subfigure}[b]{0.45\textwidth}

\begin{tikzpicture}
\filldraw[color=black, fill=black!5,very thick] (0,0) rectangle (3,0.5) ;
\filldraw[color=red!60, fill=red!5, very thick](1.5,3) circle (0.7) node[black] {$m_1$} ;
\draw[<->] (0,.6) --node[left=1pt]{$h=z_1(0)$} (0,2.2)[thin];
\filldraw[color=blue!60, fill=blue!5, very thick](1.5,5) circle (0.5)node[black]{$m_2$} ;

\draw[dashed] (0,2.3)--(2,2.3);

\draw[dashed] (1,3.75)--(3,3.75);

\draw[dashed] (1,4.5)--(3,4.5);

\draw[<->] (3,3.75) --node[right=1pt]{$\Delta h=z_2(0)-z_1(0)$} (3,4.4)[thin];

\end{tikzpicture}
\caption{Initial conditions}
\end{subfigure}
\hfill
\begin{subfigure}[b]{0.45\textwidth}
\begin{tikzpicture}
\filldraw[color=black, fill=black!5,very thick] (0,0) rectangle (3,0.5) ;
\filldraw[color=red!60, fill=red!5, very thick](1.5,3) circle (0.7) node[black] {$m_1$} ;
\draw[<->] (0,.6) --node[left=1pt]{$z_1(t)$} (0,2.2)[thin];
\filldraw[color=blue!60, fill=blue!5, very thick](1.5,5) circle (0.5)node[black]{$m_2$} ;

\draw[dashed] (0,2.3)--(2,2.3);

\draw[dashed] (1,1.9)--(3,1.9);

\draw[dashed] (1,4.5)--(3,4.5);

\filldraw[dotted, color=red!60, fill=red!5, very thick](1.5,1.2) circle (0.7) node[black] {$m_1$} ;
\draw[<->] (3,2) --node[right=1pt]{$z_2(t)$} (3,4.4)[thin];

\end{tikzpicture}
\caption{Coordinate definitions}
\end{subfigure}
\end{center}
\caption{The coordinates and initial conditions are defined to remove reference to the radii of the balls. (a) The drop height $h$ is the initial distance between the bottom of the lower ball and the floor. The drop gap $\Delta h$ is the initial distance between the top of the lower ball and the bottom of the upper ball. (b) Coordinate $z_1$ is measured from the floor to the bottom of the lower ball, and $z_2$ is measured from the top of the lower ball when it touches the floor.} \label{coords}
\end{figure}
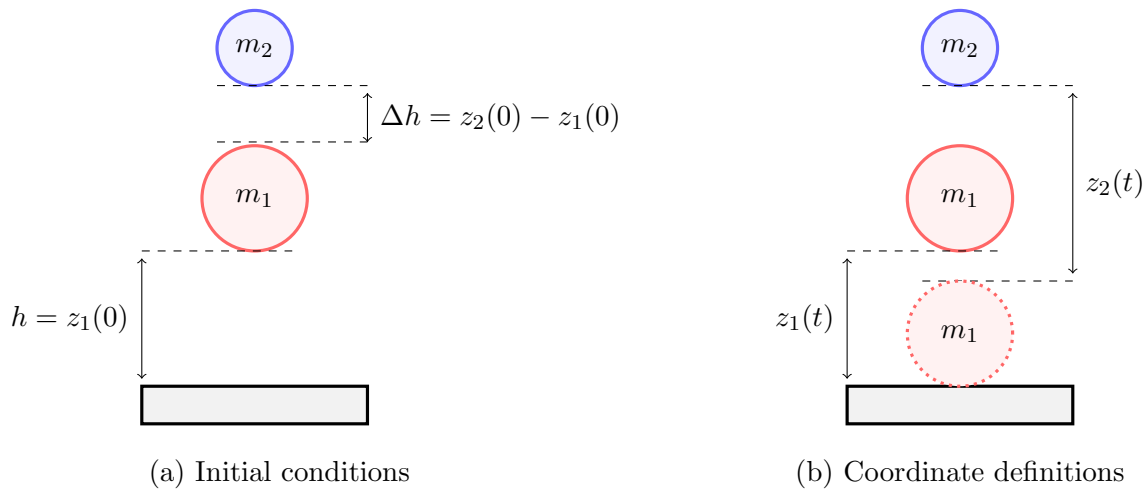

Let us consider the textbook scenario, with a tennis ball dropped on a  basketball. In one case, the lower ball is adult-sized (diameter 24 cm), and in the other case it is child-sized (diameter 20 cm). If the tennis ball achieves the same post-collision velocity in each case, the mass dropped on the basketball will bounce 4 cm higher, simply because the collision point is higher from the floor. 
To better compare these experiments, we measure the position of each ball from its lowest accessible point, which for the tennis ball is one basketball diameter above the floor.
See Fig. \ref{coords} for an illustration of this coordinate definition.
The positions $z_1=0$ and $z_2=0$ do not refer to the same physical point, removing reference to the radii of the balls.

Inelastic collisions are characterized by a coefficient of restitution, defined as the ratio of the relative velocity of the particles post-collision to their pre-collision relative velocity,
\begin{equation}
\varepsilon = \left|\frac{ v_1'-v_2'}{v_1 - v_2}\right|. \label{epsilon}
\end{equation}
Here, primed velocities refer to the velocity just after the collision.
Using this relation, along with conservation of momentum, the post-collision velocities\cite{muller_patric_two-ball_2011} are 
\begin{eqnarray}
v_1' &=& \frac{v_1+\mu \left(v_2+\varepsilon(v_2-v_1)\right)}{1+\mu} \label{postcolvel1} \\
v_2' &=& \frac{\mu v_2+v_1+\varepsilon(v_1-v_2)}{1+\mu},\label{postcolvel2}
\end{eqnarray}
where $\mu=m_2/m_1$.

To understand the first bounce, we divide the motion into five distinct phases: (i) both balls free-falling after release from rest, (ii) the lower ball colliding with the floor, (iii) both balls in free fall, with the lower ball moving upward, (iv) the collision of the two balls, (v) the upper ball in free fall until it reaches its maximum height. The collisions are approximated as instantaneous, and we assume the same coefficent of restitution $\epsilon$ for both collisions, for simplicity in the analysis.

(i) We define the drop height $h=z_1(0)$ and the gap $\Delta h=z_2(0)-z_1(0)$. The balls are released from rest simultaneously, so both fall a distance $h$ before the lower ball reaches the floor, and have a speed 
\begin{equation}
v_i=\sqrt{2gh}. \label{initialv}
\end{equation}
The upper ball has position $z_2=\Delta h$.

(ii) Using (\ref{postcolvel2}), and considering the floor to be at rest and infinitely massive, the velocity of the lower ball after colliding with the floor is $v_1=\varepsilon v_i$. 

(iii) The two balls collide when $z_1=z_2$. Using free-fall kinematics, 
\begin{eqnarray}
z_1(t) &=& v_i t-\frac{1}{2}gt^2 \\
z_2(t) &=& \Delta h -v_i t-\frac{1}{2}gt^2,
\end{eqnarray}
the collision time is 
\begin{equation}
t_c=\frac{\Delta h}{v_i(1+\varepsilon)}.
\end{equation}
This time is used to calculate the velocities of the balls just before the collision,
\begin{eqnarray}
v_1(t_c)&=&\varepsilon v_i-gt_c = \varepsilon \sqrt{2gh}-\frac{g\Delta h}{\sqrt{2gh}(1+\varepsilon)} \label{prec1} \\
v_2(t_c)  &=& -v_i-gt_c = -v_i -\frac{g\Delta h}{v_i(1+\varepsilon)} \label{prec2}
\end{eqnarray}
and the position of the collision is
\begin{equation}
z_1(t)=z_2(t)=\Delta h \frac{\varepsilon}{1+\varepsilon}-\frac{1}{4h} \left(\frac{\Delta h}{1+\varepsilon}\right)^2. \label{colheight}
\end{equation}

(iv) The post-collision velocities of the balls are calculated using (\ref{postcolvel1}-\ref{postcolvel2}), with pre-collision velocities (\ref{prec1}-\ref{prec2}). Because we are calculating the maximum height of the upper ball, we neglect the lower ball from this point on. The upper ball's velocity post-collision is
\begin{equation}
v_2'=\sqrt{2gh}\frac{\varepsilon^2+2\varepsilon-\mu}{1+\mu}-\frac{g \Delta h}{\sqrt{2gh}(1+\varepsilon)} \label{v2p}
\end{equation}

(v) We use conservation of energy to determine the maximum height of the upper ball after the first bounce 
\begin{equation}
m_2 g h_{max}=\frac{1}{2}m_2 v_2'^2 +m_2 g z_2(t_c), 
\end{equation}
Inserting (\ref{colheight}) and (\ref{v2p}) yields the expression
\begin{equation}
h_{max}=\frac{h \left( \varepsilon^2 + 2\varepsilon-\mu \right)^2 +\Delta h (1+\mu)(\mu-\epsilon) }{(1+\mu)^2}. \label{height1}
\end{equation}
This result from the ICM is used as a point of comparison for the results of the computational model in Section \ref{numerical}.

\section{Linear dashpot force} \label{hertz}
We model the inelastic collisions as a damped spring-mass system.\cite{Nagurka_2004}
The general form for this force includes a restorative force that is proportional to the compression, and a dissipative term that is proportional to the velocity
\begin{equation}
F= -kz-\gamma \dot{z}.
\end{equation} 
This model, also known as a spring with a linear dashpot, has been successfully implemented as an approximation of the Hertz contact force between viscoelastic spheres. \cite{patricio_hertz_2004, muller_patric_two-ball_2011, gugan_inelastic_2000} 
In an interaction between two such systems, effective elastic and damping constants are calculated according to
\begin{equation}
k_{ij}=\left(\frac{1}{k_i}+\frac{1}{k_j}\right)^{-1}, \qquad \gamma_{ij}=\left(\frac{1}{\gamma_i}+\frac{1}{\gamma_j}\right)^{-1}.
\end{equation}

Some changes must be made to reflect the differences between bouncing balls and damped springs.
First,  the ball only experiences a force under compression, unlike a spring which can also be stretched. To capture this, we define the compression 
$\xi_{ij}=\min[0,-z_i+z_j],$ where $z_i$ are the vertical positions of the balls as defined in Fig. \ref{coords}.
The floor is denoted by index 0, while the lower and upper balls are labeled by index 1 and 2, respectively. This quantity is zero unless the balls are in contact.

Additionally, the combination of the force terms must always result in a repulsive force. To handle cases where the dissipative term overcompensates the elastic term, resulting in an unphysical attractive force, we use the minimum function
\begin{equation}
F_{ij}=\min[0,-k \xi_{ij} -\gamma_{ij} \dot{\xi_{ij}}].\label{forcedef}
\end{equation}
To simplify the analysis, the elastic constant $k$ is taken to be the same for both forces, and the desired coefficient of restitution is set by changing the damping constants $\gamma_{ij}$.


The equations of motion are
\begin{eqnarray}
m_1 \ddot{z_1} & = & -m_1 g + F_{01} - F_{12}\\
m_2 \ddot{z_2} & = & -m_2 g + F_{12}
\end{eqnarray}
To isolate the parameters of physical importance, the equations of motion are written in a dimensionless fashion via the substitutions $z_i = X_i m_1g/k$ and $t= \tau\sqrt{m_1/k}$. 
The equations become
\begin{eqnarray}
X_1'' & = & -1 +f_{01} -f_{12} \label{EOM1}\\
X_2'' &=& -1 +f_{12}/\mu, \label{EOM2}
\end{eqnarray}
where $(')$ indicates a derivative with respect to $\tau$. The dimensionless forces are
\begin{equation}
f_{ij}=-\min\left[0,\eta_{ij}+2\zeta \eta'_{ij}\right] \label{forces}
\end{equation}
where $\zeta= \gamma_{ij}/2\sqrt{m_1k}$ are the damping ratios and $\eta_{ij}$ is the dimensionsless  compression. 

\subsection{Coefficients of restitution}

To get a general expression for the coefficient of restitution, we  solve (\ref{EOM1}-\ref{EOM2}) or the duration of the two balls, assuming no contact with the floor. Thus,  $f_{01}=0$ and $f_{12}\neq 0$, so the minimum function is unnecessary. The equations become
\begin{equation}
\eta_{ij}''=-\frac{1+\mu}{\mu}(\eta_{ij}+2\zeta \eta'_{ij}),
\end{equation}
with the initial condition $\eta_{ij}(0)=0$.  With the substitutions 
\begin{equation}
\beta=\frac{1+\mu }{\mu}\zeta, \quad \omega_0^2 =\frac{1+\mu }{\mu}, \quad \omega = \sqrt{\omega_0^2-\beta^2},
\end{equation}
the equation of motion takes on the standard damped harmonic oscillator form, with general solution 
\begin{equation}
\eta_{ij}=\frac{\eta'_{ij}(0)}{2\sqrt{\beta^2-\omega_0^2}} e^{-\beta \tau}\left( e^{\tau \sqrt{\beta^2-\omega_0^2} }-e^{-\tau\sqrt{\beta^2-\omega_0^2} } \right).
\end{equation}
In the case of low damping, $\zeta^2< \mu /(1+\mu)$, or $\beta<\omega_0$,  this becomes
\begin{equation}
\eta_{ij}=\frac{\eta'_{ij}(0)}{\omega} e^{-\beta \tau}\sin{\omega \tau}.
\end{equation}
The derivations for low damping follow. The high damping case is similar, using the definition $\Omega =\sqrt{\beta^2-\omega_0^2}$ to avoid imaginary values of $\omega$, and making use of hyperbolic trigonometric functions, beginning with the solution
\begin{equation}
\eta_{ij}=\frac{\eta'_{ij}(0)}{\Omega} e^{-\beta \tau}\sinh{\Omega \tau}.
\end{equation}


The balls cease contact at time $\eta_{ij}''(\tau_f)=0$. With low damping, this becomes 
\begin{equation}
\tan \omega \tau_f = \frac{-2\beta \omega}{\omega^2-\beta^2}.\label{trig}
\end{equation}
Solving this equation for $\tau_f$ gives an expression involving the arctangent, which requires care in selecting the correct branch. \cite{schwager_coefficient_2007} Using trigonometric definitions, the expression simplifies to
\begin{equation}
\tau_f=\frac{1}{\omega}\arccos\left(\frac{2\beta^2}{\omega_0^2}-1\right).
\end{equation}
 Converting back to physical parameters, the  contact time becomes
\begin{equation}
\tau_f = \begin{cases}
\frac{\mu }{(1+\mu)\sqrt{\frac{\mu }{1+\mu}-\zeta^2}} \arccos\left(2\zeta^2\left(\frac{1+\mu }{\mu} \right) -1\right) \quad &\mathrm{for}\, \zeta^2<\frac{\mu }{1+\mu} \\
\frac{\mu }{(1+\mu)\sqrt{\zeta^2-\frac{\mu }{1+\mu}}} \arccosh\left(2\zeta^2\left(\frac{1+\mu }{\mu} \right)-1\right) &\mathrm{for} \, \zeta^2>\frac{\mu }{1+\mu} \label{tauhigh}
\end{cases}
\end{equation}

To calculate the coefficient of restitution (\ref{epsilon}), we calculate
\begin{equation}
\varepsilon_{ij}=\left| \frac{\eta_{ij}'(\tau_f)}{\eta'_{ij}(0)}\right|=e^{-\beta \tau}\left|\cos(\omega \tau_f)-\frac{\beta}{\omega}\sin(\omega \tau_f)\right|.
\end{equation}
for low damping. Using (\ref{trig}) and trigonometric definitions, this becomes
\begin{eqnarray}
\varepsilon_{ij}&=& e^{-\beta \tau_f}\left|-\frac{\omega^2-\beta^2}{\omega_0^2}-\frac{\beta}{\omega}\frac{2\beta \omega}{\omega_0^2}\right| \\
&=& e^{-\beta \tau_f} ,
\end{eqnarray}
which also holds for high damping. Inserting (\ref{tauhigh}),  
\begin{equation}
\varepsilon_{ij} = \begin{cases}
\exp\left[\frac{\zeta } {\sqrt{\frac{\mu }{1+\mu}-\zeta^2}} \arccos\left(2\zeta^2\left(\frac{1+\mu }{\mu} \right) -1\right)\right] \quad &\mathrm{for}\, \zeta^2<\frac{\mu }{1+\mu}\\
\exp\left[\frac{\zeta } {\zeta^2-\sqrt{\frac{\mu }{1+\mu}}} \arccosh\left(2\zeta^2\left(\frac{1+\mu }{\mu} \right) -1\right)\right] &\mathrm{for} \, \zeta^2>\frac{\mu }{1+\mu}
\end{cases}
\end{equation}
The high damping case results in $\varepsilon_{ij}< e^{-2}$ in the limit $\mu \rightarrow \infty$, and even smaller for lower mass ratios. Thus, high damping is not applicable to the examples studied in Section \ref{numerical}.

\section{Numerical results from the linear dashpot model}\label{numerical}

Most analysis of the two-ball drop problem focuses on the first bounce, \cite{ee_magic_2015}  taken here to encompass the time from which the balls are released until the upper ball reaches its next local maximum in height.
The collisions of interest for the first bounce include the first collision between the lower ball and the floor, and one or several collisions between the two balls. 
We extend our analysis to include the second bounce of the upper ball. 
The lower ball  contacts the floor one or more times before the balls collide for the second time, and the second collision may or may not occur with the lower ball in contact with the floor.

The value of $\tau_f$ for the lower ball also characterizes the initial conditions of the initial drop height and gap between the balls. In the collision with the infinitely massive floor, we take the limit $\mu \rightarrow \infty$, and the contact time simplifies to
\begin{equation}
\tau_f(\mu\rightarrow \infty) = \frac{1}{\sqrt{1-\zeta^2}}\arccos(2\zeta^2-1).
\end{equation} 
The  interval between the time $X_1=0$ and $X_2=0$ is
\begin{equation}
\tau_d = \sqrt{2}\left(\sqrt{X_2(0)}-\sqrt{X_1(0)}\right).
\end{equation}
This is interpreted as the time at which the balls would collide if they did not compress or bounce. If $\tau_d < \tau_f$, the collisions are simultaneous; the lower ball is still in contact with the floor when the upper ball collides with it.

We study the maximum height of the upper ball after the first and second bounce. We analyze several different ratios: first bounce height to the ICM prediction (\ref{height1}), first and second bounce heights to the initial drop height $(h+\Delta h)$, and second bounce height to first bounce height.

For given values of $\varepsilon_{01},\, \varepsilon_{12}$, and $\mu$, the bounce height ratios are found to be the same for fixed $\tau_d/\tau_f$, and the force curves retain the same form as well. 
Thus, the situation is described entirely by three parameters ($\varepsilon_{01}, \, \varepsilon_{12}, \, \mu$) and a single initial condition ($\tau_d/\tau_f$).  
To simplify the analysis, we set $\varepsilon_{01}=\varepsilon_{12}$ throughout the rest of this work, as this does not qualitatively affect the results. \cite{muller_patric_two-ball_2011}

The ball parameters $\varepsilon,\mu$ and the initial condition $\tau_d/\tau_f$ were each  divided into 50 increments, covering ranges $\varepsilon \in (0.5,1)$, $\mu \in (10^{-2}, 1/3)$ and $\tau_d/\tau_f \in (10^{-2},1)$ for analysis of the bounce heights.
For the sake of comparison, the plots shown here focus primarily on the representative value of $\varepsilon=0.816$, but animated visualizations of all data are available. \cite{bartz_data_nodate}
We use a standard numerical ordinary differential equation solver\cite{hindmarsh_ODEPACK_1983, Petzold_Automatic_1983} to integrate the initial value problem (\ref{EOM1}, \ref{EOM2}) with the balls released from rest. 

\subsection{Details of first bounce collisions}

\begin{figure}
\centering
\includegraphics[width=\columnwidth]{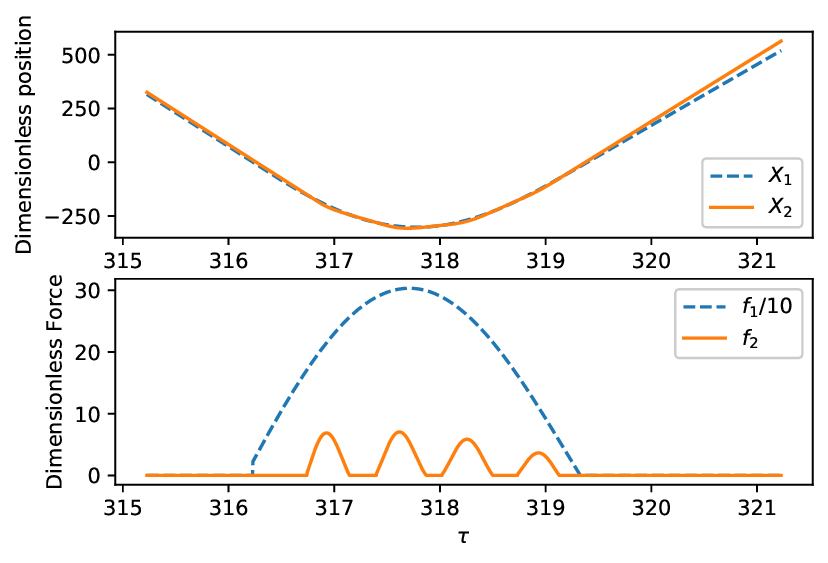}
\caption{The trajectories of the balls and the force curves during the first collision are shown for a representative case where $\tau_d/\tau_f=0.010.$ In this case, $\mu=0.010$, and $\varepsilon_{01}=\varepsilon_{12}=0.9$. }
\label{closeup}
\end{figure}

Analysis of the forces \eqref{forces} from simulations with simultaneous collisions reveals multiple contacts between the two balls under a variety of conditions, consistent with previous analysis. \cite{muller_patric_two-ball_2011}
Figure \ref{closeup} shows a case with four contacts between the balls, occurring entirely while the lower ball is in contact with the floor.
The compression of the balls is illustrated by  $X_1$ and $X_2$ becoming negative in Fig. \ref{closeup}. 

\begin{figure}[htb]
\centering
\includegraphics[width=\columnwidth]{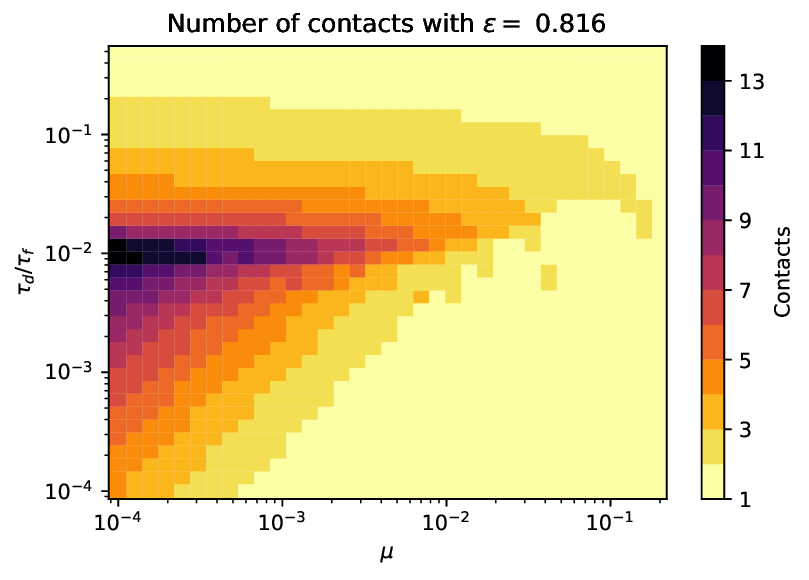}
\caption{The number of contacts between the two balls during the first bounce with a typical coefficient of restitution $\varepsilon=0.816$, chosen to be the same for both balls. The axes are logarithmically scaled to illustrate the effects found with small initial gaps and mass ratios.}\label{contactcount}
\end{figure}

\begin{figure}[htb]
\centering
\includegraphics[width=\columnwidth]{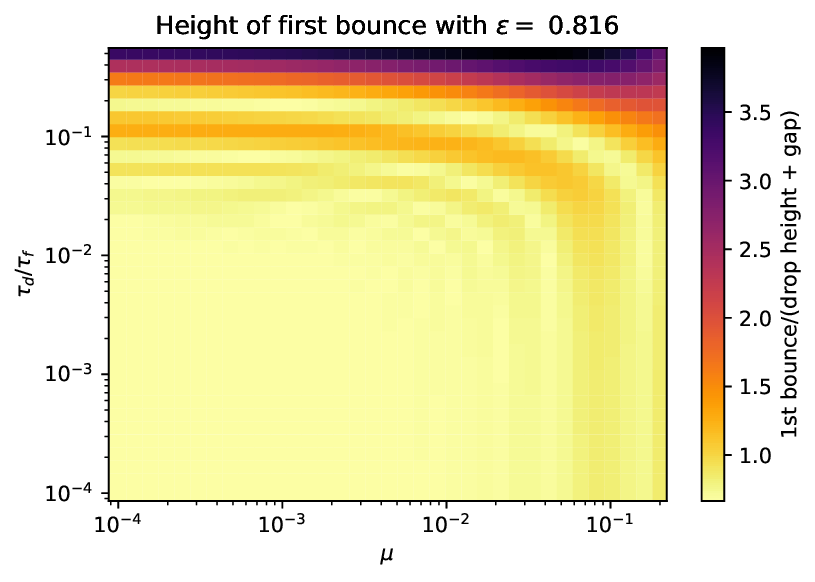} 
\caption{Comparison to Fig. \ref{contactcount} shows no clear relationship between number of contacts and the height the upper ball reaches on its first bounce. \label{bounceICMlog}}
\end{figure}

The number of contacts between the two balls depends sensitively upon the ball parameters and initial condition. 
The count varies between one and twelve for the representative data shown in Fig. \ref{contactcount}, with $\varepsilon=0.816$.
Multiple contacts are mostly found when $\mu$ and $\tau_d/\tau_f$ are small, so the plot axes are scaled logarithmically to show this detail, and the range of $\tau_d/\tau_f$ is expanded to ($10^{-4},1$). Comparison to Fig. \ref{bounceICMlog} shows that the number of contacts between the balls does not correlate with the bounce height of the upper ball.
 
%

Because contact between the two balls is defined by $f_2 \neq 0$, instead of by the balls' positions, experimental setups that measure position only cannot confirm or refute the details of the collisions presented here. \cite{berdeni_two-ball_2015}
Piezoelectric sensors placed between the balls and on the floor present a more promising experimental setup, but the situations studied  do not show clear evidence for multiple contacts.  \cite{cross_vertical_2007}
However, these particular measurements do not conflict with our simulations, as they do not match conditions predicted to exhibit multiple collisions.
Future experiments targeting the parameters that predict multiple contacts  would help validate the use of this model.

\subsection{Comparison of first bounce heights to ICM}
The post-collision motion of the balls is more readily observable in an experimental setting than the forces between the balls. 
We focus on the maximum height of the upper ball after each bounce rather than the relative velocity of the balls because 
the height of the second bounce is influenced by both the post-collision velocity and the height of the second collision.

\begin{figure}[htb]
\centering
\includegraphics[width=\columnwidth]{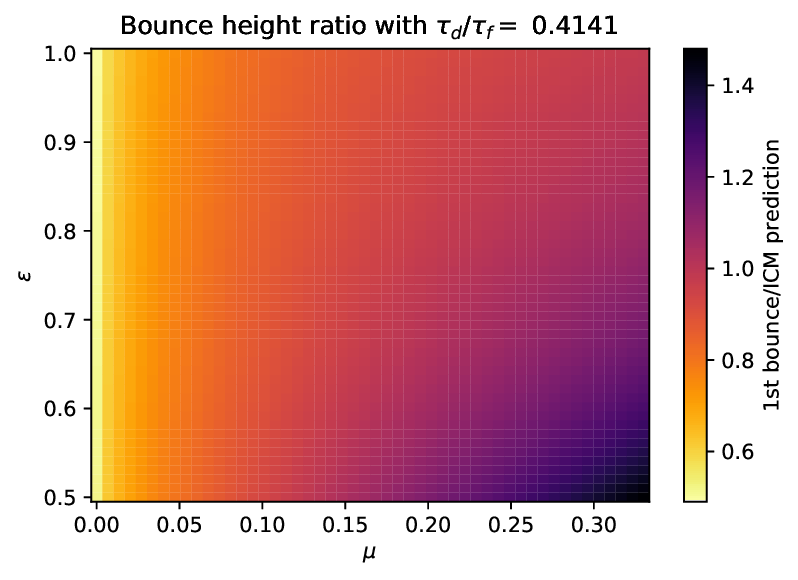}
\caption{For moderate initial gaps, as shown here, there are some combinations of $\varepsilon, \, \mu$ that result in bounces lower than the ICM prediction, while others result in bounces higher than the ICM. The ratios plotted range from 0.49 to 1.48.}
\label{belowICM}
\end{figure}

When the initial gap between the balls is small ($\tau_d/\tau_f\ll1$), the height of the first bounce is less than the ICM prediction \eqref{height1}. 
The ICM limit is recovered in the case $\tau_d/\tau_f > 1$, as expected when collisions are independent.
Intermediate  gaps show more detailed structure. 
For some combinations of parameters, the bounce height remains below the ICM prediction, while for others the simulated bounce height is greater than that of the ICM.
The ICM is overperformed when $\varepsilon$ is small and $\mu$ is on the large end of the range studied, as shown in Fig. \ref{belowICM}. 
These conditions lead to small bounce heights, below the initial drop height, in any case.

\begin{figure}[htb]
\centering
\includegraphics[width=\columnwidth]{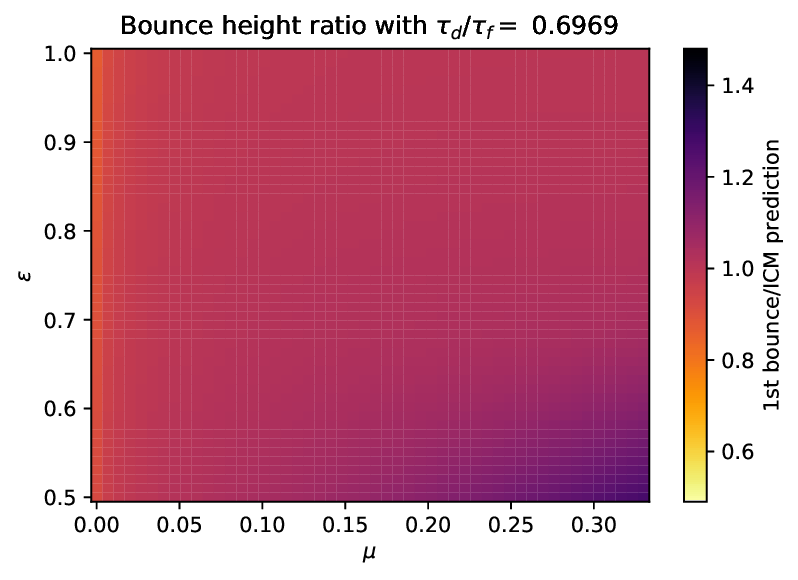}
\caption{The results of the simulation begin to to converge to the ICM result for a wide range of $\varepsilon, \, \mu$, despite the collisions not being independent, with $\tau_d/\tau_f<1$. The ratios plotted range from 0.87 to 1.27.}
\label{nearICM}
\end{figure}

The ICM closely approximates post-collision behavior for a variety of cases where the collisions are not truly independent, as $\tau_d/\tau_f<1$. 
For $\tau_d/\tau_f \approx 0.7$, there is deviation from the ICM for some regions of the parameter space shown in Fig. \ref{nearICM}, but most situations are well-approximated by the simple model.
Plots of larger values of $\tau_d/\tau_f $ are not shown because they do not exhibit noticeable contrast, as all results are quite close to the ICM value. 
For example, the bounce heights are all within 5\% of the ICM limit for $\tau_d/\tau_f>0.85$ for the range of ball parameters studied.
This unexpected success of the ICM is confirmed by experimental measurements. \cite{berdeni_two-ball_2015, cross_vertical_2007}

\subsection{Second bounce} \label{bounce2}

\begin{figure}[htb]
\centering
\includegraphics[width=\columnwidth]{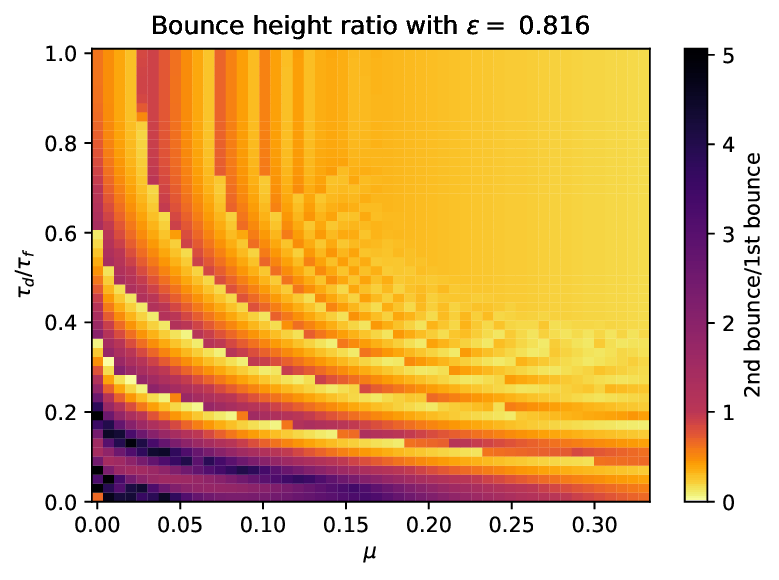}
\caption{Comparison of the second bounce height to the first bounce height. Red points indicate situations in which the second bounce is higher than the first.} \label{2ndbounce}
\end{figure}

In this section, we analyze the height of the upper ball on the second bounce, specifically studying the delayed rebound effect with the second bounce higher than the first.
Figure \ref{2ndbounce} shows a comparison of the second bounce height to the height of the first bounce using $\varepsilon=0.816$ as an illustrative example.
This plot shows that most cases of the delayed rebound effect occur for small initial drop gaps and small mass ratios.
The region expands as $\varepsilon$ increases.
As expected, the small mass-ratio limits are approximately achieved in the elastic case ($\varepsilon=1$), when the initial collisions between the floor and the two balls are independent ($\tau_d/\tau_f=1$). 

\begin{figure*}[htb]
\centering
\subfloat[]{
\includegraphics[width=0.49\textwidth]{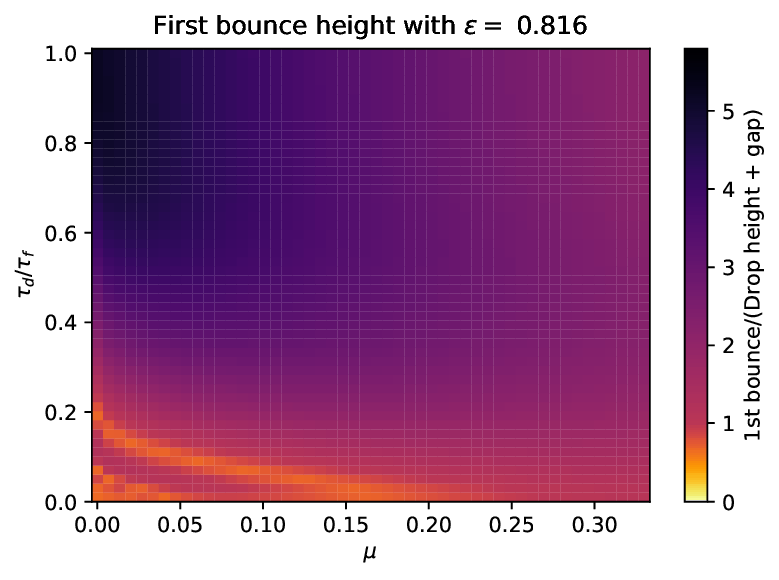}}
\subfloat[]{\includegraphics[width=0.49\textwidth]{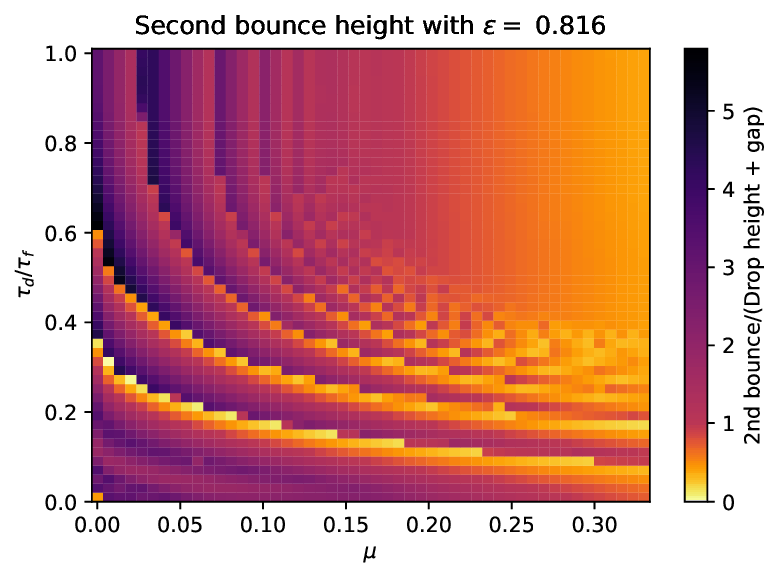}}
\caption{Comparison of bounce heights to the initial height for a representative value of the coefficient of restitution. The plot in (a) shows that the first bounce exceeds the drop height in most regions of parameter space. In (b), we see more detailed structure for the second bounce.  
\label{bounce12}
 }
\end{figure*}

In Fig. \ref{bounce12}, the first and second bounce heights are compared. 
It is evident that the delayed rebound effect is most prominent in cases where the first bounce is low, often lower than the initial drop height.
The highest second bounces on an absolute scale occur when the first bounce is also high. 
Some of these do slightly exceed the first bounce, but these cases are relatively few.

The general sense of the delayed rebound effect obtained by visual inspection  is confirmed by systematic comparison.
For the parameter ranges studied, 12.3\% of combinations resulted in the delayed rebound effect. 
Of these, 93\% occurred in cases where the first bounce was lower than the ICM prediction. 
However, having a first bounce lower than the ICM prediction was not highly predictive; only 22\% of these cases have a higher second bounce.

Typically, the lower ball is not in contact with the floor when the balls collide for the second bounce, and the balls make a single contact with each other.
When the lower ball is in contact with the floor, multiple contacts between the balls are possible, in a manner that is qualitatively similar to the first bounce.
These cases lead to a noticeably lower second bounce than others in nearby parameter space.

\section{Conclusions}

We show that the ``delayed rebound effect," where the second bounce of two aligned balls is higher than the first, is present in the numerical solutions to a linear dashpot force between the balls.
The effect is is most prominent when the upper ball has a much smaller mass than the lower ball, and the distance gap between the balls is small when they are released.
Typically, the finite duration of the collisions leads the first bounce to be smaller than predicted by the ICM, so the expected high bounce does not come until the second bounce.
This is consistent with the informal observation that inspired this study: namely, that the delayed rebound effect is more readily produced in ping-pong ball--basketball collisions than in tennis ball--basketball collisions.

Rather than focus on the parameters of a few particular sports balls, we look for universal behavior. 
The relevant ball parameters are reduced to the mass ratio $\mu=m_1/m_2$ and the coefficient of restitution $\varepsilon$, assumed to be the same for both balls. 
The initial conditions of drop height and gap between the balls is reduced to a single parameter $\tau_d/\tau_f$, which characterizes the time between the lower ball reaching the floor and the two balls making first contact. 
This approach represents a simplification over previous studies, and can be generalized to more than two balls.

We examined the details of the first bounce collisions, finding multiple contacts between the balls in some cases.
However, the multiple impacts did not correlate with the post-collision dynamics of the balls.
For the first bounce, we found that small initial gaps resulted in lower bounces   than the ICM prediction, while the ICM is approximately correct for larger initial gaps, despite the overlapping collisions. 

There are several extensions of this work that can serve as student research projects.
This universal study can be applied to realistic ball parameters for both first and second bounces.
Detailed study of two-particle interactions can also be extended to chain collisions of multiple particles in a line. \cite{kerwin_velocity_1972, patricio_hertz_2004,redner_2004, kires_astroblasterfascinating_2009, gharib_shock_2011, ricardo_maximizing_2015} or studied in the chaotic r\'egime. \cite{whelan_1990}
While the linear dashpot force gives qualitatively similar results to the Hertz force on the first bounce, \cite{muller_patric_two-ball_2011}  simulating these forces in chain collisions yields qualitatively different results, \cite{hinch_fragmentation_1999} so an interested student may wish to extend this study to such a force. 

This toy model can serve as a student's introduction to impact mechanics, which is relevant in a variety of fields including engineering, granular materials, and molecular dynamics. 
The results of this model may be relevant to study of inelastic collapse and clumping in granular flows, \cite{mcnamara_1992, constantin_1995, zhou_1996, luding_1999, topic_inelastic_2019} particularly in the presence of gravity or another driving force. \cite{wakou_2013}
Inelastic collapse involves an infinite number of collisions in finite time, which presents challenges to event-driven modeling, \cite{reichardt_event_2007} so understanding when the ICM is accurate can help improve simulation efficiency. 

A thorough experimental study of the delayed rebound effect requires a mechanism to constrain the interacting particles to a single dimension. 
Precisely aligned spheres, as used in experimental studies of the first bounce, are unlikely to remain aligned for a second bounce.
Low friction carts on an inclined track, with springs for repulsion, may be a useful model, although it might be difficult to achieve the range of mass ratios seen in ball drop experiments.

\bibliography{main}

\begin{thebibliography}{10}

\bibitem{mellen_superball_1968}
Walter~Roy Mellen.
\newblock Superball {Rebound} {Projectiles}.
\newblock {\em American Journal of Physics}, 36(9):845--845, September 1968.

\bibitem{harter_velocity_1971}
William~G. Harter.
\newblock Velocity {Amplification} in {Collision} {Experiments} {Involving}
  {Superballs}.
\newblock {\em American Journal of Physics}, 39(6):656--663, June 1971.

\bibitem{herrmann_simple_1981}
F.~Herrmann and P.~Schmälzle.
\newblock Simple explanation of a well-known collision experiment.
\newblock {\em American Journal of Physics}, 49(8):761--764, August 1981.

\bibitem{cross_vertical_2007}
Rod Cross.
\newblock Vertical bounce of two vertically aligned balls.
\newblock {\em American Journal of Physics}, 75(11):1009--1016, October 2007.

\bibitem{berdeni_two-ball_2015}
Y.~Berdeni, A.~Champneys, and R.~Szalai.
\newblock The two-ball bounce problem.
\newblock {\em Proc. R. Soc. Lond. A}, 471(2179):20150286, July 2015.

\bibitem{muller_patric_two-ball_2011}
{Muller, P.} and T.~Poschel.
\newblock Two-ball problem revisited: {Limitations} of event-driven modeling.
\newblock {\em Physical Review E}, 83(4):041304, April 2011.

\bibitem{Nagurka_2004}
M.~Nagurka and Shuguang Huang.
\newblock A mass-spring-damper model of a bouncing ball.
\newblock In {\em Proceedings of the 2004 American Control Conference},
  volume~1, pages 499--504 vol.1, 2004.

\bibitem{patricio_hertz_2004}
P.~Patr\'icio.
\newblock The {Hertz} contact in chain elastic collisions.
\newblock {\em American Journal of Physics}, 72(12):1488--1491, November 2004.

\bibitem{gugan_inelastic_2000}
D.~Gugan.
\newblock Inelastic collision and the {Hertz} theory of impact.
\newblock {\em American Journal of Physics}, 68(10):920--924, September 2000.

\bibitem{schwager_coefficient_2007}
Thomas Schwager and Thorsten Pöschel.
\newblock Coefficient of restitution and linear–dashpot model revisited.
\newblock {\em Granular Matter}, 9(6):465--469, November 2007.

\bibitem{ee_magic_2015}
June-Haak Ee and Jungil Lee.
\newblock Magic mass ratios of complete energy-momentum transfer in
  one-dimensional elastic three-body collisions.
\newblock {\em American Journal of Physics}, 83(2):110--120, January 2015.

\bibitem{bartz_data_nodate}
Sean Bartz.
\newblock Data and visualizations.

\bibitem{hindmarsh_ODEPACK_1983}
A.~C. Hindmarsh.
\newblock {ODEPACK}, {A} {Systematized} {Collection} of {ODE} {Solvers}.
\newblock {\em IMACS Transactions on Scientific Computation}, 1:55--64, 1983.

\bibitem{Petzold_Automatic_1983}
L.~R. Petzold.
\newblock Automatic selection of methods for solving stiff and nonstiff systems
  of ordinary differential equations.
\newblock {\em SIAM Journal on Scientific and Statistical Computing},
  4(1):136--148, 1983.

\bibitem{kerwin_velocity_1972}
James~D. Kerwin.
\newblock Velocity, {Momentum}, and {Energy} {Transmissions} in {Chain}
  {Collisions}.
\newblock {\em American Journal of Physics}, 40(8):1152--1158, August 1972.

\bibitem{redner_2004}
S.~Redner.
\newblock A billiard-theoretic approach to elementary one-dimensional elastic
  collisions.
\newblock {\em American Journal of Physics}, 72(12):1492--1498, 2004.

\bibitem{kires_astroblasterfascinating_2009}
Marián Kireš.
\newblock Astroblaster--a fascinating game of multi-ball collisions.
\newblock {\em Physics Education}, 44(2):159--164, February 2009.

\bibitem{gharib_shock_2011}
Mohamed Gharib, Ahmet Celik, and Yildirim Hurmuzlu.
\newblock Shock {Absorption} {Using} {Linear} {Particle} {Chains} {With}
  {Multiple} {Impacts}.
\newblock {\em Journal of Applied Mechanics}, 78(3):031005, May 2011.

\bibitem{ricardo_maximizing_2015}
Bernard Ricardo and Paul Lee.
\newblock Maximizing kinetic energy transfer in one-dimensional many-body
  collisions.
\newblock {\em European Journal of Physics}, 36(2):025013, February 2015.

\bibitem{whelan_1990}
N.~D. Whelan, D.~A. Goodings, and J.~K. Cannizzo.
\newblock Two balls in one dimension with gravity.
\newblock {\em Phys. Rev. A}, 42:742--754, Jul 1990.

\bibitem{hinch_fragmentation_1999}
E.~J. Hinch and S.~Saint-Jean.
\newblock The fragmentation of a line of balls by an impact.
\newblock {\em Proc. R. Soc. Lond. A}, 455(1989):3201--3220, September 1999.

\bibitem{mcnamara_1992}
Sean McNamara and W.~R. Young.
\newblock Inelastic collapse and clumping in a one‐dimensional granular
  medium.
\newblock {\em Physics of Fluids A: Fluid Dynamics}, 4(3):496--504, 1992.

\bibitem{constantin_1995}
Peter Constantin, Elizabeth Grossman, and Muhittin Mungan.
\newblock Inelastic collisions of three particles on a line as a
  two-dimensional billiard.
\newblock {\em Physica D: Nonlinear Phenomena}, 83(4):409 -- 420, 1995.

\bibitem{zhou_1996}
Tong Zhou and Leo~P. Kadanoff.
\newblock Inelastic collapse of three particles.
\newblock {\em Phys. Rev. E}, 54:623--628, Jul 1996.

\bibitem{luding_1999}
S.~Luding and H.~J. Herrmann.
\newblock Cluster-growth in freely cooling granular media.
\newblock {\em Chaos: An Interdisciplinary Journal of Nonlinear Science},
  9(3):673--681, 1999.

\bibitem{topic_inelastic_2019}
Nikola Topic and Thorsten Pöschel.
\newblock Inelastic collapse of perfectly inelastic particles.
\newblock {\em Communications Physics}, 2(1):85, July 2019.

\bibitem{wakou_2013}
Jun'ichi Wakou, Hiroyuki Kitagishi, Takahiro Sakaue, and Hiizu Nakanishi.
\newblock Inelastic collapse in one-dimensional driven systems under gravity.
\newblock {\em Phys. Rev. E}, 87:042201, Apr 2013.

\bibitem{reichardt_event_2007}
Roland Reichardt and Wolfgang Wiechert.
\newblock Event driven algorithms applied to a high energy ball mill
  simulation.
\newblock {\em Granular Matter}, 9(3):251--266, June 2007.

\end{thebibliography}
\bibliographystyle{unsrt}

\end{document}